# Terrestrial Test of the Shapiro Time Delay, Fourth Test of Einstein's General Theory of Relativity

(July 4th, 2026)

Farhad Hakimi and Hosain Hakimi, emails: fhakimi@alum.mit.edu, hhakimi@alum.mit.edu

**Abstract:**

The Shapiro time delay is one of the four classical tests of general relativity, yet all previous experimental measurements have relied on astronomical observations involving planetary radar ranging, spacecraft tracking, or astrophysical systems. We propose a compact fiber-optic Sagnac interferometer that enables a terrestrial laboratory measurement of the Shapiro time delay using Earth's gravitational field. The proposed instrument exploits the differential gravitational propagation delay between vertically separated fiber coils and converts this extremely small relativistic effect into a measurable optical phase shift through reciprocal interferometric detection, thereby establishing coherent optical phase as a new observable for Shapiro-delay measurements alongside the traditional observables of propagation time, range, and Doppler tracking. The approach provides a direct experimental pathway toward determining the first post-Newtonian 1PN parametrized post-Newtonian PPN parameter gamma, while establishing a scalable framework for future investigations of second post-Newtonian 2PN propagation effects through measurement of the second-order coefficient C, which depends on the PPN parameters beta, gamma, and epsilon. We develop the theoretical formulation of the terrestrial Shapiro delay, describe the interferometer architecture, and analyze the expected signal, dominant noise sources, and achievable sensitivity using realistic photonic and electronic components. The analysis indicates that the proposed approach is compatible with current fiber-optic sensing technology and advanced digital signal-processing techniques. By translating a classical astronomical test of general relativity into a compact laboratory experiment, this work introduces a scalable platform for precision measurements of relativistic light propagation and establishes a new experimental methodology for testing gravitational physics under controlled terrestrial conditions.

## I. Introduction

The Shapiro time delay is one of the four classical tests of general relativity, together with the perihelion advance of Mercury, gravitational light deflection, and gravitational redshift. It was first proposed by Irwin I. Shapiro in the 1960s during radar astronomy experiments in which electromagnetic signals were transmitted to planets and returned to Earth [1-3]. Shapiro recognized that signals passing through regions of strong gravitational potential, particularly near the Sun during planetary superior conjunctions, experience an additional propagation delay arising from spacetime curvature. This effect, now known as the *Shapiro time delay*, provided a direct experimental confirmation that gravity influences the propagation of electromagnetic waves.

Subsequent measurements using planetary radar ranging, spacecraft tracking, very long baseline interferometry, and binary pulsar observations have confirmed the relativistic prediction with increasing precision. The most precise direct determination of the Shapiro delay was obtained from the Cassini spacecraft experiment [4], which constrained the first post-Newtonian parametrized post-Newtonian (PPN) parameter ($\gamma$) at the level of a few parts in ($10^5$). Despite these remarkable achievements, all existing measurements rely on astronomical or interplanetary propagation paths and therefore require favorable celestial alignments, long baselines, or dedicated space missions.

A controlled terrestrial measurement of the Shapiro time delay would provide a complementary approach, enabling repeatable laboratory investigations of relativistic light propagation. Previous terrestrial proposals have explored the use of large-scale gravitational-wave interferometers combined with externally generated time-varying gravitational fields [5,6]. These approaches, however, require extremely sensitive kilometer-scale observatories and are ultimately limited by the strength and characterization of laboratory source masses.

In this work, we introduce a different approach based on a reciprocal fiber-optic Sagnac interferometer that measures the static gravitational potential gradient of Earth itself. The proposed architecture employs vertically separated fiber coils with nearly identical optical paths, allowing the differential gravitational propagation delay to be extracted while suppressing common-mode disturbances through reciprocal interferometric detection. The long optical interaction length provided by coiled fiber, combined with coherent modulation, sensitive photodetection, and digital signal processing, enables a practical pathway toward laboratory measurement of the terrestrial Shapiro effect.

The primary objective of this work is to establish the feasibility of determining the first post-Newtonian PPN parameter ($\gamma$) through a terrestrial interferometric experiment. The same platform also provides a foundation for future investigations of higher-order post-Newtonian effects involving the PPN parameters. We develop the theoretical formulation of the terrestrial Shapiro delay, analyze the proposed fiber-optic Sagnac sensor, evaluate signal and noise performance using realistic implementation parameters, and present a differential measurement protocol designed to suppress uncertainties.

The history of Shapiro-delay measurements illustrates how advances in experimental technology continually open new regimes for testing general relativity. The original planetary-radar experiments measured microsecond-scale propagation delays over astronomical distances. The present work explores the opposite extreme: detecting sub-attosecond-equivalent gravitational propagation effects within a compact terrestrial interferometer by exploiting the phase sensitivity of coherent optical measurements.

The principal contributions of this work are:
(1) the introduction of a compact reciprocal fiber-optic Sagnac architecture for terrestrial measurement of the Shapiro time delay;
(2) the development of a laboratory-scale formulation connecting differential gravitational propagation delay to measurable optical phase shifts and PPN parameters; and
(3) a comprehensive analysis of sensor sensitivity, noise limitations, digital signal processing, and experimental implementation strategies for precision tests of relativistic gravity.

## II. Astronomical Formulation of the Shapiro Time Delay

In general relativity, the propagation time of an electromagnetic signal is increased when it traverses a gravitational field. To first post-Newtonian (1PN) order, the one-way Shapiro time delay for a signal propagating in the field of a spherical gravitating body is given by [7].

$$\delta t = \frac{2GM}{c^3} \ln\left(\frac{4r_1 r_2}{d^2}\right). \qquad (1)$$

where (G) is Newton's gravitational constant, (M) is the mass of the gravitating body, (c) is the speed of light, ($r_1$) and ($r_2$) are the distances of the emitter and receiver from the gravitating body, and (d) is the minimum distance between the signal path and the center of the mass (Fig. 1). Equation (1) corresponds to the near-opposition approximation, in which the emitter, gravitating body, and receiver are approximately aligned and the logarithmic contribution to the delay is maximized.

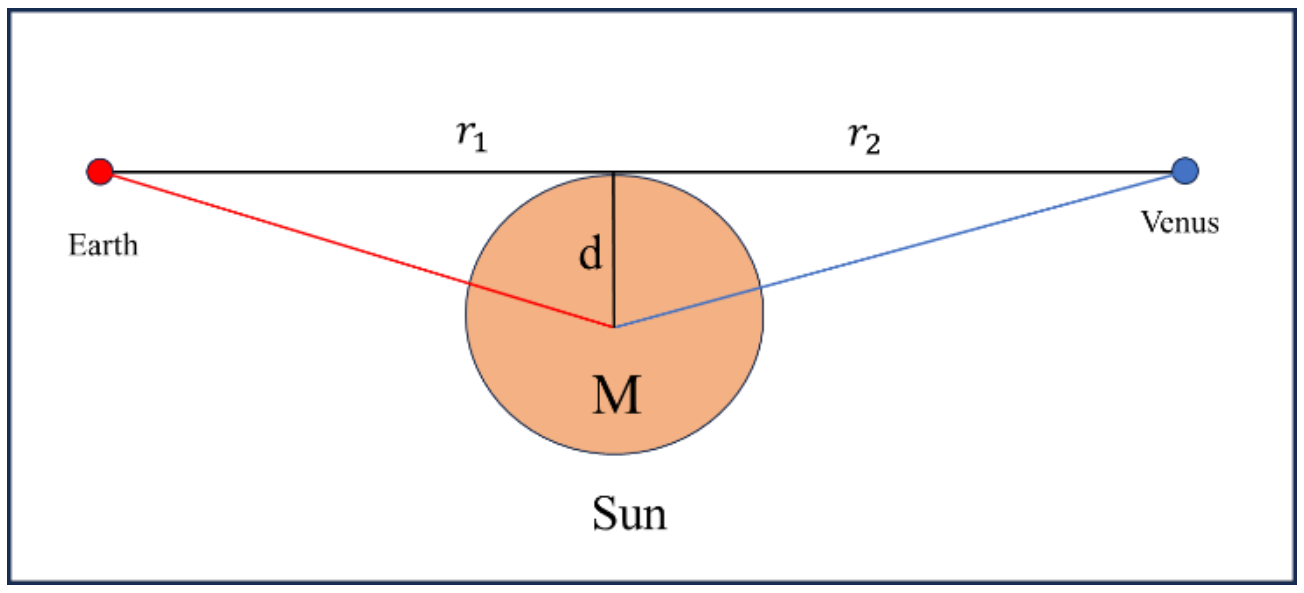

**Figure 1:** Shapiro spacetime delay experiment: the fourth classical test of general relativity.

The classical solar-system measurements of the Shapiro time delay were performed using radar signals transmitted to planets and spacecraft during superior conjunctions, when the signal path passed close to the Sun. The measured excess propagation time agreed with the predictions of general relativity and established the Shapiro effect as the fourth classical test of Einstein's theory.

Within the parameterized post-Newtonian (PPN) framework [8,9], deviations from general relativity are described by dimensionless parameters that characterize possible modifications of weak-field gravity. To first post-Newtonian order, the Shapiro delay depends on the PPN parameter (γ), which describes the amount of spatial curvature produced per unit mass. The generalized expression is therefore

$$\delta t = (1+\gamma)\frac{GM}{c^3}\ln\left(\frac{4r_1 r_2}{d^2}\right) \tag{2}$$

General relativity predicts (γ=1), recovering the standard Shapiro time-delay expression.

The most precise direct measurement of the Shapiro delay to date was obtained from the Cassini-Huygens spacecraft experiment during a solar conjunction, which constrained deviations from general relativity to the level of $(\gamma - 1) = (2.1 \pm 2.3) \times 10^{-5}$, in agreement with the relativistic prediction [4]. Other constraints on (γ) have been obtained from planetary radar ranging, lunar laser ranging, very long baseline interferometry (VLBI), and analyses of planetary dynamics.

All previous experimental measurements of the Shapiro time delay have relied on astronomical or interplanetary propagation paths. A terrestrial experiment therefore requires a different formulation because the optical path remains close to Earth's surface and the gravitational potential varies only weakly over laboratory dimensions.

Previous terrestrial proposals have explored the use of gravitational-wave interferometers combined with externally generated source masses. For example, Ballmer and collaborators proposed measuring a time-varying gravitational Shapiro delay using advanced interferometric gravitational-wave detectors [5,6]. Their approach demonstrated the possibility of laboratory tests of relativistic light propagation, but requires large-scale observatories and relies on extremely small gravitational perturbations produced by laboratory masses.

The approach developed in this work instead exploits Earth's gravitational field directly using a compact fiber-optic Sagnac interferometer. The following sections derive the corresponding terrestrial Shapiro delay and develop the experimental architecture for measuring the resulting differential propagation-time shift.

## III. Terrestrial Shapiro Delay for Laboratory Measurements

### A. Calculation on Earth Surface:

According to general relativity, electromagnetic propagation through a gravitational field experience an increase in coordinate travel time due to spacetime curvature. This effect, known as the Shapiro time delay, does not imply a local reduction of the speed of light, which remains equal to $c$ for all local observers. Instead, the additional propagation time arises from the geometry of spacetime.

For a static, spherically symmetric gravitational field, such as the Schwarzschild field of the Earth, light propagation in the weak-field limit can be represented using an equivalent optical-medium description. In this formulation, spacetime curvature is expressed through an effective position-dependent refractive index.

To first post-Newtonian (1PN) order, the effective refractive index is

$$n = 1 - \frac{(1+\gamma)\,\Phi}{c^2}, \qquad (3)$$

where $\Phi$ is the Newtonian gravitational potential and $\gamma$ is the PPN parameter describing the amount of spatial curvature produced per unit mass.

Since the Earth's gravitational potential is $\Phi = -\frac{GM_E}{r}$. Equation (3) becomes

$$n = 1 + \frac{(1+\gamma)GM_E}{c^2 r}, \qquad (4)$$

where $M_E$ is the mass of Earth and $r$ is the distance from the Earth's center.

Although this effective refractive-index description does not represent a physical optical medium, it provides an equivalent first-order representation of light propagation in curved spacetime. The propagation time is obtained from the optical path integral

$$T = \frac{1}{c}\int n\, ds, \qquad (5)$$

where $ds$ is the Euclidean path-length element. To 1PN order, this formulation reproduces the propagation time obtained directly from the Schwarzschild metric.

Consider two fiber-optic coils, each having optical length $L_{\text{fiber}}$, positioned at different elevations above the Earth's surface (Fig. 2). The coils form the two differential optical paths of the Sagnac interferometer described in the following section. Because the two coils experience slightly different gravitational potentials, their effective refractive indices are different, producing a differential Shapiro propagation delay.

For two coils located at radii $R_E$ and $R_E + d$, the difference in refractive index is

$$\begin{aligned} \Delta n &= n_{\text{low}} - n_{\text{high}} \\ &= \frac{(1+\gamma)\,GM_E}{c^2}\left(\frac{1}{R_E} - \frac{1}{R_E + d}\right). \end{aligned} \qquad (6)$$

For $d \ll R_E$, this becomes

$$\Delta n \simeq \frac{(1+\gamma)\, GM_E}{c^2} \frac{d}{R_E^2}. \qquad (7)$$

The corresponding first post-Newtonian terrestrial Shapiro time-delay difference is therefore

$$\Delta t_{1PN}^{\text{Shapiro}} = \frac{\Delta n\, L_{\text{fiber}}}{c} \qquad (8)$$

or

$$\Delta t_{1PN}^{\text{Shapiro}} = (1+\gamma) \frac{GM_E}{c^3} \frac{d\, L_{\text{fiber}}}{R_E^2} \qquad (9)$$

For general relativity, γ=1.

Equation (9) represents the terrestrial analogue of the astronomical Shapiro delay. Instead of relying on a long signal path passing near a massive body, the proposed experiment exploits the small gravitational-potential difference between two vertically separated optical paths while accumulating the effect over a long fiber length.

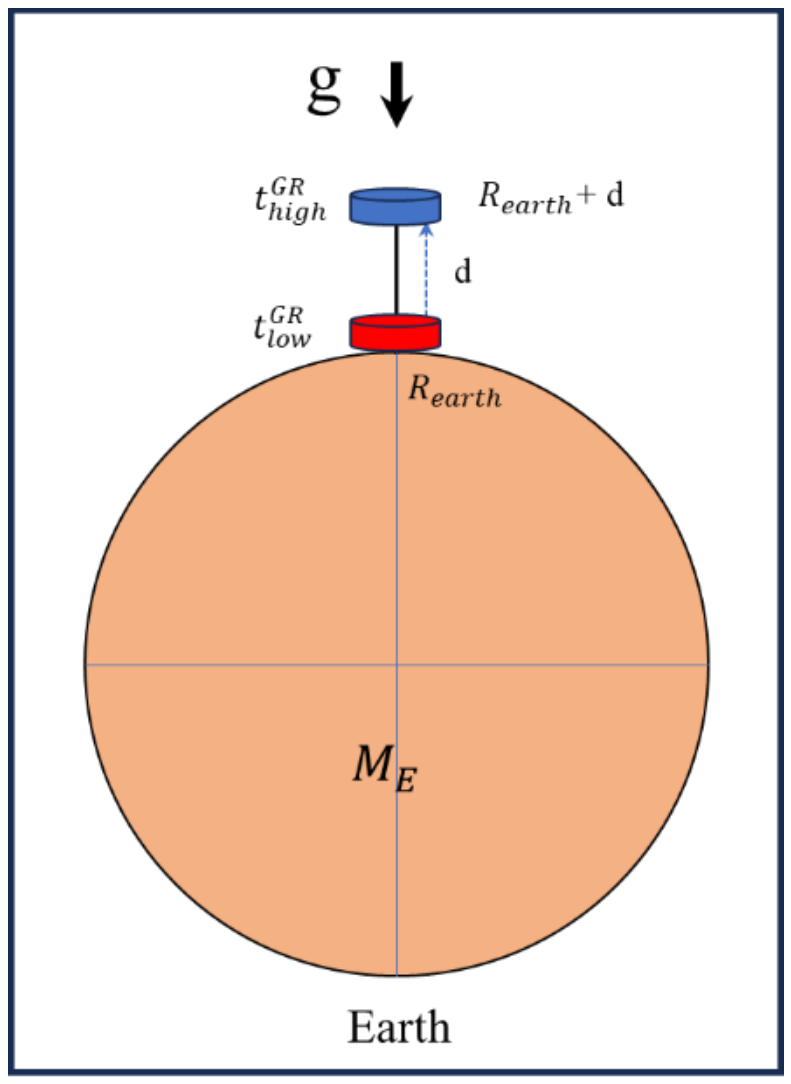

**Figure 2:** Schematic illustration of the Shapiro time delay for upper and lower fiber-optic coils separated by a vertical distance $d$ on Earth's surface (not to scale).

**B. Second Post-Newtonian (2PN) Shapiro Time-Delay Correction:** The first post-Newtonian (1PN) contribution depends exclusively on the PPN parameter ($\gamma$). Beyond 1PN order, nonlinear gravitational effects are described by the coefficient $C$, a specific combination of the PPN parameters ($\beta$), ($\gamma$), and ($\epsilon$) that encapsulates the higher-order relativistic corrections.

Following the second post-Newtonian (2PN) PPN expansion of Epstein and Shapiro [10] and subsequent treatments of relativistic light propagation, the effective optical refractive index may be written as

$$n = 1 - \frac{(1+\gamma)\Phi}{c^2} + C\left(\frac{\Phi}{c^2}\right)^2 + \cdots , \qquad (10)$$

where

$$C = \frac{3}{2} - \beta + \gamma - \frac{1}{2}\gamma^2 + \frac{3}{4}\epsilon. \qquad (11)$$

Using the optical path integral and retaining the second-order term gives the differential 2PN Shapiro contribution

$$\Delta t_{2PN}^{\text{Shapiro}} = 2C\,\frac{G^2 M_E^2\, L_{\text{fiber}}\, d}{c^5\, R_E^3}. \qquad (12)$$

Substituting the expression for $C$,

$$\Delta t_{2PN}^{\text{Shapiro}} = \left(3 - 2\beta + 2\gamma - \gamma^2 + \frac{3}{2}\epsilon\right)\frac{G^2 M_E^2\, L_{\text{fiber}}\, d}{c^5 R_E^3} \qquad (13)$$

For general relativity,

$$\gamma = \beta = \epsilon = 1,$$

and therefore

$$t_{2PN}^{\text{Shapiro}} = \frac{7}{2}\frac{G^2 M_E^2\, L_{\text{fiber}}\, d}{c^5 R_E^3} \qquad (14)$$

The ratio of the 2PN correction to the leading 1PN contribution is

$$\frac{\Delta t_{2PN}^{\text{Shapiro}}}{\Delta t_{1PN}^{\text{Shapiro}}} = \frac{7}{4}\frac{G M_E}{R_E\, c^2}. \qquad (15)$$

For Earth,

$$\frac{\Delta t_{2PN}^{\text{Shapiro}}}{\Delta t_{1PN}^{\text{Shapiro}}} \approx 1.2 \times 10^{-9} \qquad (16)$$

Thus, the second-order correction is approximately nine orders of magnitude smaller than the leading 1PN signal. Although experimentally challenging, measurement of this contribution would provide sensitivity to the nonlinear structure of gravity and to the higher-order PPN parameters ($\beta$) and ($\epsilon$).

## IV. Fiber-Optic Sagnac Sensor Architecture

### A. Motivation

The terrestrial Shapiro time delay predicted by Eq. (9) is extremely small, requiring a measurement technique capable of detecting ultrashort differential propagation-time shifts. Conventional fiber-optic timing systems do not possess sufficient sensitivity for this purpose. However, the combination of long optical interaction lengths, reciprocal interferometric detection, and differential gravitational-potential sampling provides a practical approach for amplifying and measuring this relativistic effect.

The proposed sensor is based on a reciprocal fiber-optic Sagnac interferometer, a technology originally developed for high-precision rotation sensing. In contrast to conventional interferometric fiber-optic gyroscopes (IFOGs), which measure the nonreciprocal Sagnac phase produced by rotation (Fig. 4), the present architecture is designed to detect the reciprocal gravitational propagation-time difference between two optical paths located at slightly different gravitational potentials.

The use of long coiled optical fibers enables a large effective interaction length within a compact laboratory footprint. By comparing two nearly identical fiber paths separated vertically by a distance $d$, the interferometer converts the extremely small gravitational time-delay difference into a measurable optical phase shift while rejecting common-mode environmental disturbances.

### B. Geometry of the Fiber-Loop Sensor

The proposed sensor consists of two fiber loops wound on cylindrical spools, each containing approximately $L_{\text{fiber}} \approx 100$ km of single-mode optical fiber (Fig. 3). The spools are arranged to form the two arms of a reciprocal Sagnac interferometer. One spool serves as the reference path, while the second spool can be displaced vertically by a distance $d$. When both coils are located at the same elevation, they experience essentially identical gravitational potentials and provide a calibration reference. When one coil is displaced vertically, the resulting potential difference produces the terrestrial Shapiro time-delay signal derived in Section III.

The differential measurement therefore depends not on the absolute gravitational potential, but on the local gravitational-potential gradient. This differential configuration suppresses sensitivity to common environmental effects while isolating the relativistic propagation-time contribution.

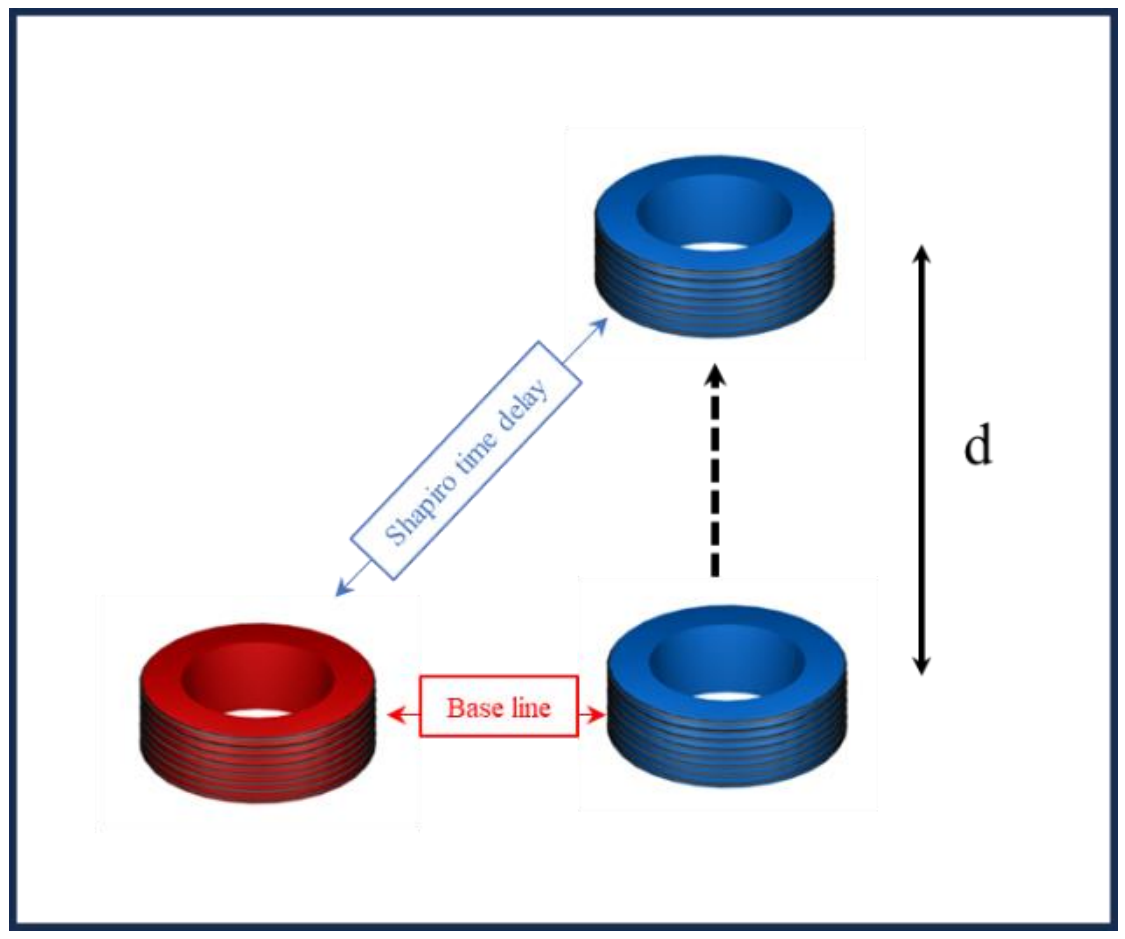

**Figure 3:** Schematic illustration of two fiber spools, each containing approximately 100 km of single-mode optical fiber. One spool (shown in red) remains stationary, while the second is initially positioned at the same elevation for calibration and subsequently raised to a vertical height $d$ to measure the Shapiro time delay.

### C. Numerical Estimate

To illustrate the expected signal magnitude, consider a representative implementation with $L_{\text{fiber}} = 100$ km and a vertical displacement $d = 1$ m. Using Eq. (9), the predicted first post-Newtonian terrestrial Shapiro delay is $\Delta t_{1PN}^{\text{Shapiro}} = 7.2 \times 10^{-20}$ s. The equivalent optical path-length difference is $c\, \Delta t_{1PN}^{\text{Shapiro}} \simeq 2.2 \times 10^{-11}$ m $= 22$ pm.

Although this displacement is extremely small, it is within the range of modern interferometric measurements when combined with coherent optical detection, narrowband modulation, long integration times, advanced digital signal processing.

The corresponding 2PN contribution is $\Delta t_{2PN}^{\text{Shapiro}} \simeq 6 \times 10^{-29}$ s, which is approximately nine orders of magnitude smaller than the 1PN signal. Detecting this higher-order contribution would require further improvements in instrumental sensitivity and error control, but the same interferometric platform provides a natural framework for such future measurements.

## D. Reciprocal Sagnac Operation and Common-Mode Rejection

The exceptional sensitivity of the proposed instrument is inspired by the remarkable performance of interferometric fiber-optic gyroscopes (IFOGs), which routinely measure extremely small nonreciprocal optical phase shifts while suppressing environmental disturbances through reciprocal propagation (Fig. 4). State-of-the-art IFOGs have demonstrated the ability to resolve optical path-length differences approaching the scale of a proton diameter [11,12].

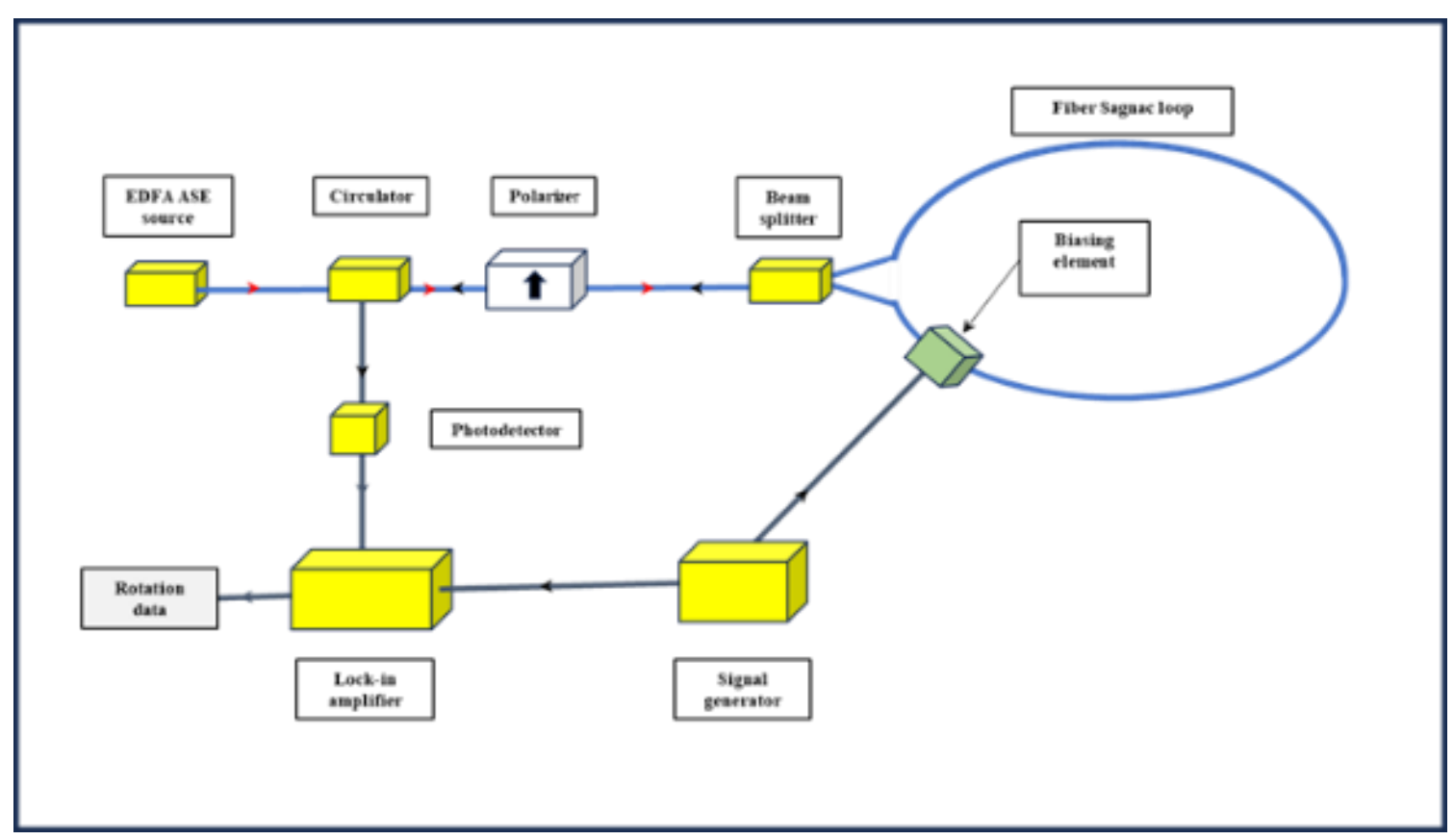

**Figure 4**: Interferometric Fiber Optic Gyroscope (IFOG) minimum-configuration architecture.

Like an interferometric fiber-optic gyroscope, the proposed sensor employs a reciprocal fiber-optic Sagnac interferometer, in which clockwise and counterclockwise optical waves traverse the same fiber path in opposite directions (Fig. 5). Because both waves experience nearly identical environmental perturbations—including temperature fluctuations, mechanical stress, vibration, acoustic disturbances, and fiber inhomogeneities—these common-mode effects are largely canceled through reciprocity. Consequently, the residual phase shift is dominated by the differential gravitational propagation delay between the upper and lower fiber coils.

This operating principle closely resembles the minimum-configuration interferometric fiber-optic gyroscope (Fig. 4), whose near-perfect reciprocity has enabled some of the highest sensitivities achieved in fiber-optic interferometry [11,12]. The proposed instrument therefore combines the maturity, stability, and extraordinary sensitivity of modern fiber-optic gyroscope technology with a fundamentally new application, establishing a practical laboratory platform for precision measurements of relativistic gravitational time delay and future experimental tests of post-Newtonian gravity, including independent determinations of the PPN parameters $\gamma$, $\beta$, and $\varepsilon$.

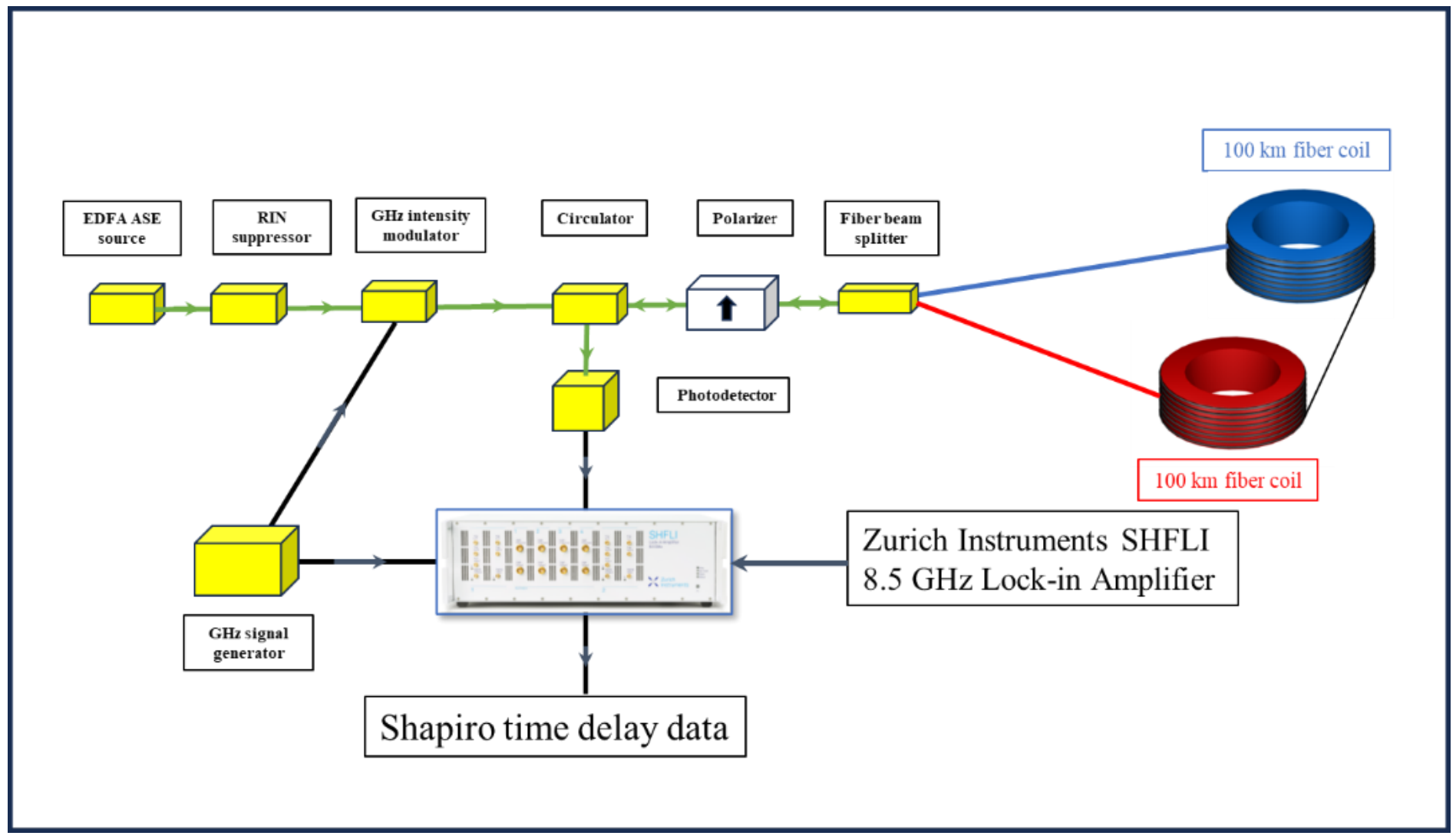

Figure 5: General architecture of the Sagnac Shapiro time delay sensor.

It is noteworthy that the quantity $c\,\Delta t_{\text{Shapiro}}$ contains no dependence on the fiber's refractive index. In the Sagnac configuration, the time delay arises purely from spacetime geometry and is independent of the propagation medium—an effect well established in relativistic analyses of Sagnac interferometers [13 and references therein].

The paired-loop geometry is specifically designed to maximize sensitivity to the vertical gravitational-potential gradient of Earth's gravitational field while simultaneously rejecting non-gravitational perturbations. As illustrated in Figs. 3 and 5, the vertically separated fiber coils sample slightly different gravitational potentials, producing the differential Shapiro time delay that constitutes the desired measurement signal. Because the measurement depends on the vertical gravitational-potential gradient rather than the absolute gravitational potential, common-mode environmental effects are inherently suppressed.

## V. Interferometer Performance: Signal and Noise Analysis

Assessing the sensitivity of a fiber-based Sagnac interferometer for measuring the Shapiro time delay requires a quantitative comparison between the expected signal response and the cumulative noise contributions from the optical source, fiber propagation, and detection electronics. The achievable performance is conveniently characterized by the resulting signal-to-noise ratio (SNR), which we analyze below.

### A. Shapiro Time-Delay Signal Response

An interferometer biased at the quadrature point, as shown in Fig. 6, Shapiro delay signal is described by Eqs. (17)–(20). Here, $P_{detector}$ is the total optical power received by the detector. The biasing element is set to 90° (the quadrature point), placing the interferometer at its point of maximum slope for optimal detection sensitivity.

$$\phi^{\text{Shapiro}} = \frac{2\pi}{\lambda_0} c\, \Delta t^{Shapiro} \quad ; \tag{17}$$

$$\phi^{Bias} = \frac{\pi}{2} \; ; \tag{18}$$

$$P_{detector} = \frac{P_{\text{opt}}}{2} \left(1 + \cos\left(\phi^{Bias} + \phi^{\text{Shapiro}}\right)\right) = \frac{P_{\text{opt}}}{2} \left(1 - \sin\left(\phi^{\text{Shapiro}}\right)\right); \tag{19}$$

The signal is encoded as a change in optical power produced by the phase shift induced by the Shapiro time delay.

$$\frac{\partial P_{detector}}{\partial \phi_{Shapiro}} = \frac{P_{\text{opt}}}{2} \left(-cos\left(\phi^{\text{Shapiro}}\right)\right) \cong -\frac{P_{\text{opt}}}{2} \tag{20}.$$

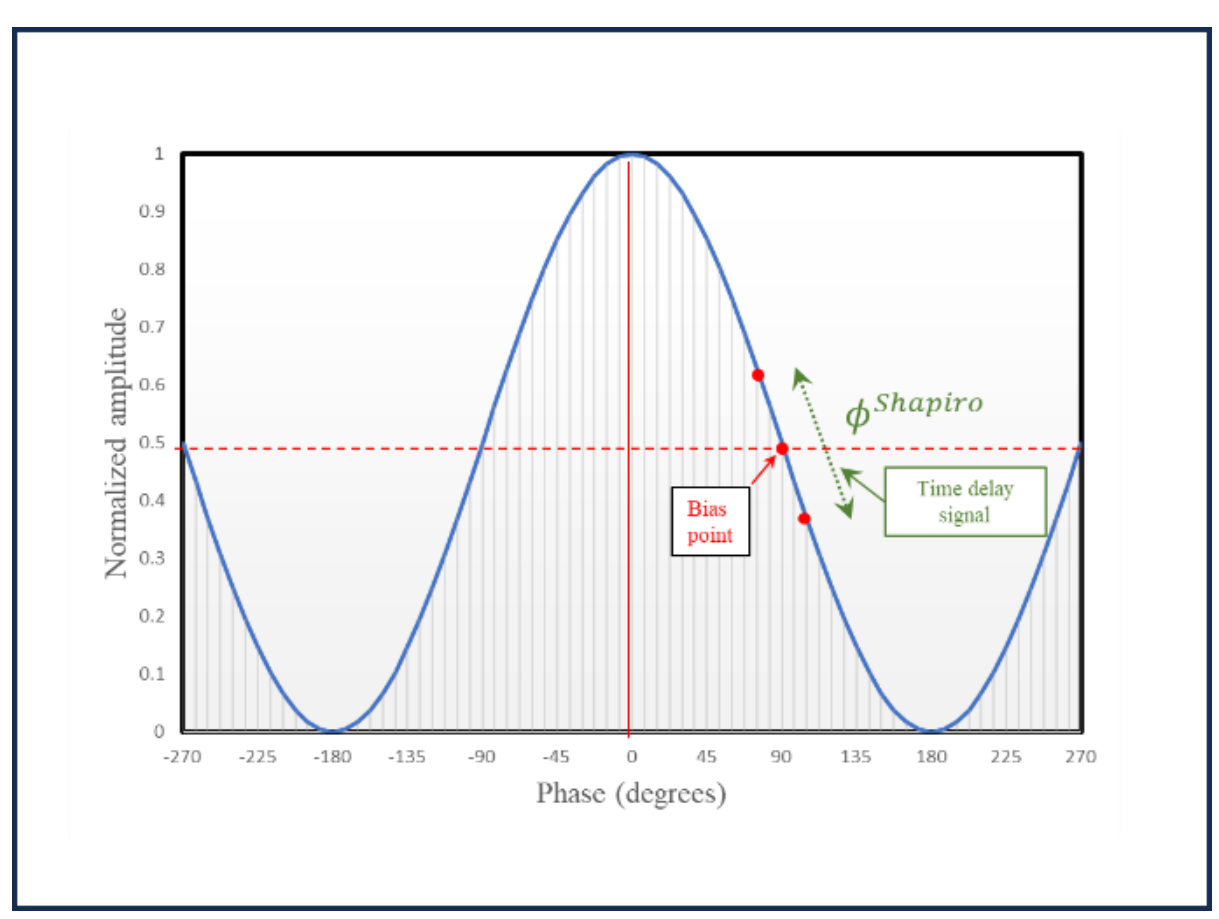

**Figure 6:** The biasing element sets the interferometer at quadrature—its point of maximum slope—to achieve optimal sensitivity.

**B. Noise in Optical Source:** In a Sagnac interferometer using broadband light, relative intensity noise (RIN)—random optical-power fluctuations—is often the dominant noise source, exceeding shot noise unless actively suppressed.

A practical mitigation strategy uses a two-stage source. Broadband light is first generated from amplified spontaneous emission (ASE) in an EDFA. Although spectrally broad and relatively flat, ASE exhibits high RIN. The ASE output is then sent through a semiconductor optical amplifier (SOA) driven deep into saturation, where gain depletion forces the SOA to act as a power equalizer that suppresses intensity fluctuations.

Commercial SOAs operated in this regime can achieve up to 22 dB of RIN suppression, roughly a hundredfold reduction in noise power [14]. The main limitation is bandwidth: suppression is typically effective only up to ~4 GHz, restricting how much of the RIN spectrum can be mitigated.

**C. Noises in Fiber:** Techniques developed for fiber-optic gyroscopes apply directly to Shapiro time-delay fiber sensor, especially the use of broadband, short-coherence sources that suppress nonlinear effects such as self- and cross-phase modulation. Two fundamental fiber noise sources, however, remain: thermo-optic noise, arising from temperature-induced changes in refractive index and length; and thermomechanical noise, small longitudinal length fluctuations driven by thermal vibrations and mechanical loss. In balanced interferometers such as the Sagnac, thermomechanical noise is strongly suppressed by common-mode rejection.

Wanser's seminal work *"Fundamental Phase Noise Limit in Optical Fibers Due to Temperature Fluctuations"* [15] provides the governing formulas. His analysis shows that both noise mechanisms fall rapidly with increasing frequency and decreasing temperature—making high-frequency operation and cryogenic cooling essential for a Sagnac-based Shapiro delay detector.

Figure 7 illustrates thermo-optic noise versus frequency at several temperatures. For example, a 200 km single-mode fiber cooled to the superfluid helium regime (~2 K) reaches phase-noise levels near –100 dB/√Hz at ~1.8 GHz, far below the room-temperature or low-frequency limits. Furthermore, the exceptionally high thermal conductivity of superfluid helium also ensures a nearly uniform temperature along the entire fiber, further reducing thermally induced index fluctuations.

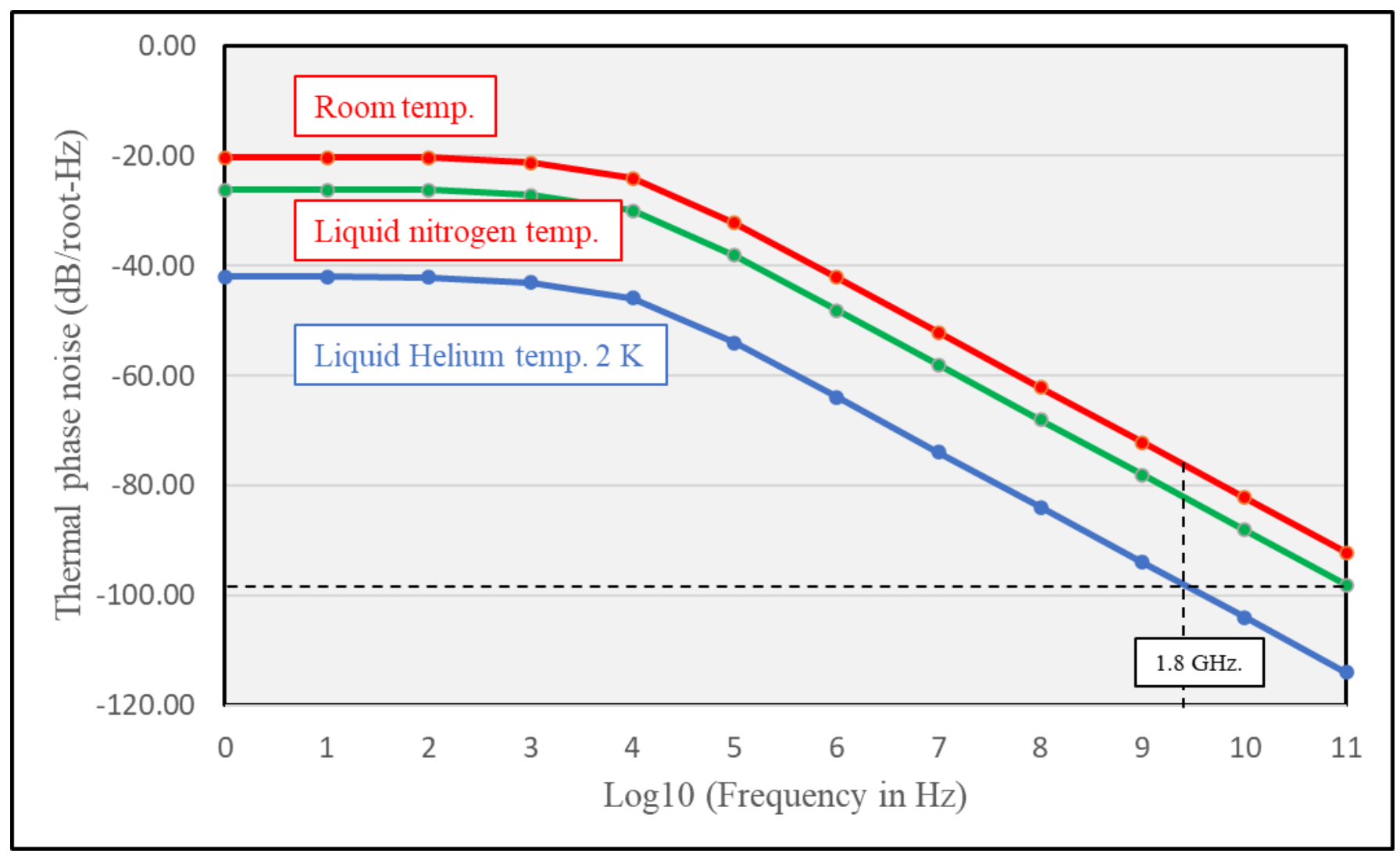

**Figure 7** : Phase noise of a 200 km single-mode fiber as a function of frequency at three different temperatures. The plot highlights the advantage of operating at higher frequencies and lower temperatures to reduce phase noise.

**D. Noise Sources in the Detection System:** The detection system is limited by three fundamental noise mechanisms: optical shot noise, optical RIN, and electronic noise. In the expressions that follow, $h\nu$ is the photon energy, $P_{detector}$ is the total optical power received by the detector, $\lambda_0$ is the mean optical wavelength, NEP is the detector noise-equivalent power, $R$ is the RIN-suppression factor, $\mathrm{BW_{elec}}$ is the electronic bandwidth, and $\Delta\nu_{\mathrm{optical}}$ is the optical bandwidth. Expressions (20) to (22) give the quantitative contributions of each noise source.

$$Shot\ noise = \sqrt{\left(\frac{2hc}{\lambda_0}\right)\ (P_{detector})\ BW_{elec}}\ ; \qquad (21)$$

$$RIN_{opt} = \frac{1}{R}\ (P_{detector})\ \sqrt{\frac{BW_{elec.}}{\Delta\nu_{opt}}}\ ; \qquad (22)$$

$$Elec\ noise = NEP\ \sqrt{BW_{elec}}\ . \qquad (23)$$

The equations above define noise contributions for the Shapiro time delay—shot noise, electronic noise, and relative intensity noise (RIN), as summarized below:

$$\Delta t_{shot\ noise}^{Shapiro} = (\frac{1}{c}\ \frac{\lambda_0}{2\pi})\ \sqrt{(\frac{2hc}{\lambda_0})\ \frac{BW_{elec}}{P_{detector}}}\ \ ; \qquad (24)$$

$$\Delta t_{RIN}^{Shapiro} = \left(\frac{1}{c}\ \frac{\lambda_0}{2\pi}\right)\frac{1}{R}\ \sqrt{\frac{BW_{elec.}}{\Delta\nu_{opt}}}\ \ . \qquad (25)$$

$$\Delta t_{elec\ noise}^{Shapiro} = \left(\frac{1}{c}\ \frac{\lambda_0}{2\pi}\right)\ \frac{NEP\ \sqrt{\ BW_{elec}}}{P_{detector}}\ \ ; \qquad (26)$$

## VI. Representative Implementation and Design Parameters

Let us consider an example of the proposed pair-loop Sagnac detector optimized for measuring the Shapiro time delay. To illustrate the design, we assign representative parameters to the configuration: each coil consists of 100 km of single-mode optical fiber wound with zero effective enclosed area. The coils are arranged in a Sagnac interferometer configuration with vanishing net area, thereby rendering the system insensitive to rotation-induced Sagnac phase shifts. The biasing element is set to 90°, placing the interferometer at its point of maximum response slope and thus maximizing detection sensitivity, as illustrated in Fig. 6.

The depolarized amplified spontaneous emission (ASE) output from the erbium-doped fiber amplifier (EDFA) is routed through a semiconductor optical amplifier (SOA) to suppress relative intensity noise (RIN), and is then intensity-modulated at 1.8 GHz, a frequency selected to remain below the SOA bandwidth limit. Assuming 300 mW [16] of input power is launched into the 200 km fiber sensor, and the fiber attenuation is 0.15 dB/km, approximately 0.3 mW reaches to a high-speed photodiode.

Figure 8 illustrates the impact of different noise sources on the Sagnac Shapiro time delay detector. The plot shows that relative intensity noise (RIN) dominates the sensitivity limit unless it is suppressed. With effective RIN reduction, the detector can reach its targeted sensitivity.

The values to generate Fig. 8 plot are, wavelength 1.5 μm, optical bandwidth of 30 nm ($3.75\text{x}10^{12}$ Hz), 13 dB RIN-suppression, electronic bandwidth of 1 Hz, and 0.3 mW optical received by the high-speed photo-detector with NEP of $2\text{x}10^{-12}$ W/√Hz.

The figure indicates that a delay detection sensitivity of approximately $\Delta t \sim 10^{-23}$ s is achievable even prior to any post–data processing. Since the expected first post-Newtonian (1PN) Shapiro time delay is approximately $\Delta t_{1PN}^{Shapiro} = 7.2 \times 10^{-20}$ s the proposed sensor possesses sufficient intrinsic sensitivity to detect the 1PN terrestrial Shapiro effect with a comfortable measurement margin. Expressed in terms of the parametrized post-Newtonian (PPN) parameter (**γ)**, this instrumental timing sensitivity corresponds to a nominal statistical sensitivity of approximately $\delta\gamma \approx \Delta t/\Delta t_{1PN}^{Shapiro} = 10^{-23}/7.2 \times 10^{-20} \approx 1.4 \times 10^{-4}$ assuming all other experimental parameters are known exactly. In practice, the ultimate uncertainty in γ will also depend on the accuracy of Earth's gravitational parameter, geometric calibration of the interferometer, and environmental noise.

As discussed in the following section, coherent integration and digital post-processing techniques can further improve the effective timing sensitivity by several orders of magnitude, thereby substantially increasing the achievable measurement precision for the Shapiro time delay and the corresponding PPN parameters.

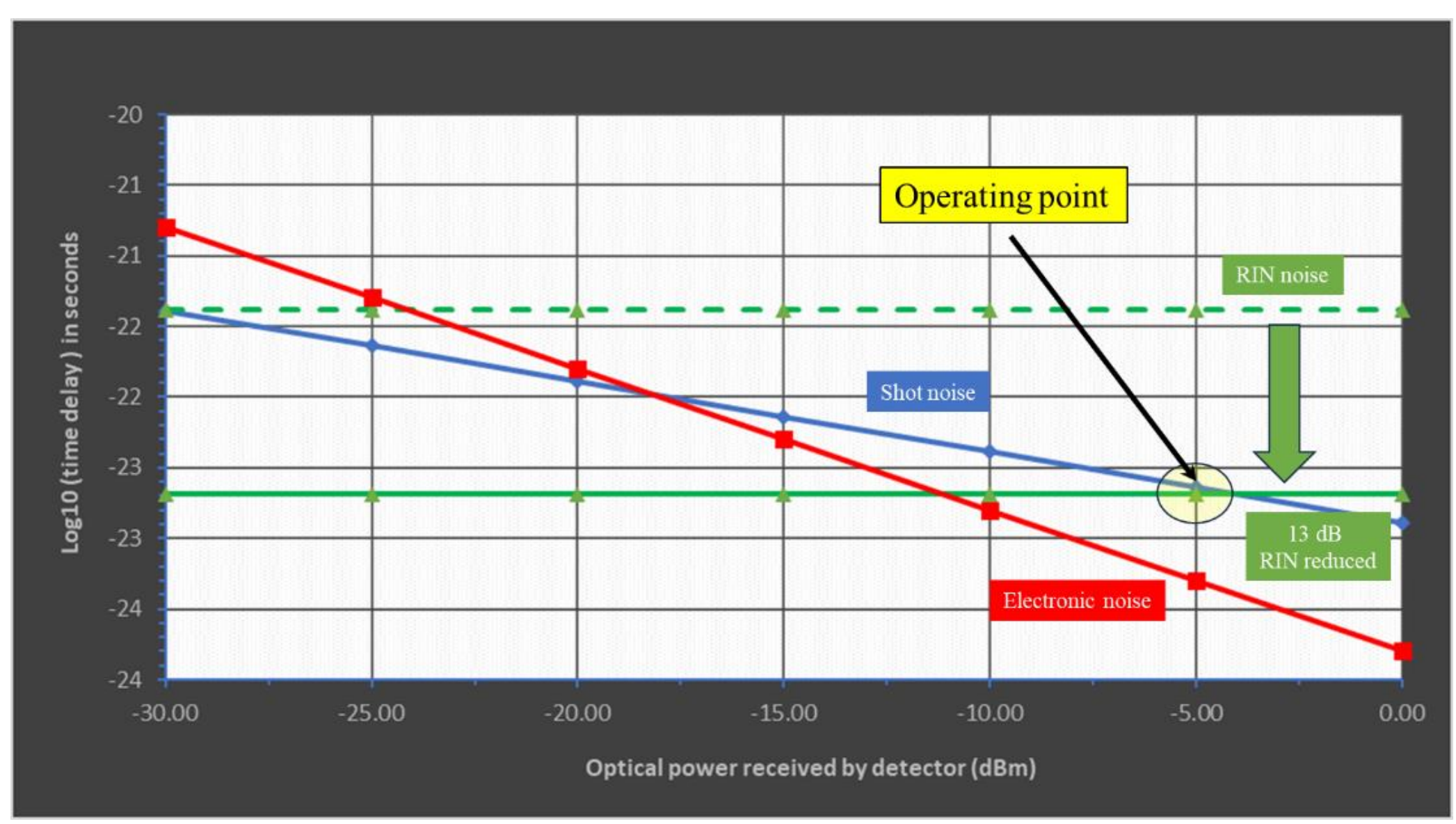

**Figure 8:** Illustrates the various noise contributions affecting sensor sensitivity, including relative intensity noise (RIN) before and after suppression. A raw (pre-processed) delay time sensitivity of approximately $10^{-23}$ seconds can be achieved at a received power of -5 dBm by a photodiode.

Notably, to avoid the overwhelming low-frequency noise inherent in long fibers, the Sagnac interferometer is operated at a high frequency (GHz). The low-frequency time signals (Hz) are then extracted through demodulation with a lock-in amplifier, effectively shifting the measurement to a cleaner spectral region while preserving the signal of interest. This approach demonstrates how careful frequency management can exploit the interferometer's sensitivity while mitigating practical noise limitations, providing a clear illustration of both the physics and engineering behind the design.

## VII. Signal Post-Processing for SNR Enhancement

### A. Enhancement factor due to oversampling

The Shapiro time delay measured by the proposed interferometer is extracted as a slowly varying (approximately DC to 1 Hz) signal following coherent modulation and demodulation. For such narrowband signals, digital post-processing can substantially improve the signal-to-noise ratio (SNR) without increasing either the optical power or the measurement bandwidth. The overall processing gain results from the combined effects of coherent demodulation, oversampling, narrowband digital filtering, coherent averaging, and long integration times.

Oversampling is particularly effective because it spreads quantization noise over a much wider frequency range than that occupied by the desired signal [17]. Subsequent digital filtering retains only the narrow signal band, thereby reducing the in-band quantization noise. For white quantization noise, the theoretical SNR improvement is determined by the oversampling ratio (OSR), defined as

$$\mathrm{OSR} = \frac{f_s}{2\,B} \tag{27}$$

where $f_s$ is the ADC sampling frequency and B is the signal bandwidth. For a commercial 14-bit analog-to-digital converter (ADC) operating at $f_s$= 6 GSa/s and a demodulated signal bandwidth of B=1 Hz the oversampling ratio becomes

$$\mathrm{OSR} = \frac{6 \times 10^9}{2 \times 10^0} = 3 \times 10^9. \tag{28}$$

This corresponds to an ideal oversampling gain of

$$10\ log_{10}(OSR) = 95dB \tag{29}$$

The achievable improvement is also limited by the intrinsic quantization performance of the ADC. For a 14-bit converter, the quantization-limited signal-to-noise ratio is approximately

$$\mathrm{SNR}_{\mathrm{ADC}} = 6.02 \times 14 + 1.76 \approx 86\ \mathrm{dB}. \tag{30}$$

Consequently, the theoretical upper limit to the digital processing gain is

$$\min(86\ \mathrm{dB}, 95\ \mathrm{dB}) = 86\ \mathrm{dB}. \tag{31}$$

In practice, however, the achievable improvement is reduced by nonideal effects, including finite effective number of bits (ENOB), thermal noise, sampling-clock jitter, analog front-end imperfections, and other system nonlinearities. Taking these practical limitations into account, a conservative overall digital processing gain of approximately 75 dB is adopted throughout this work. This value should be interpreted as an engineering estimate rather than a fundamental limit.

The predicted noise performance shown in Fig. 8 indicates that the proposed interferometer achieves a raw (pre-processed) differential time-delay sensitivity of approximately $\delta \approx 10^{-23}\ s$ within a 1 Hz measurement bandwidth. Applying a representative digital processing improvement of 75 dB, corresponding to an amplitude improvement factor of approximately $10^{75/20} \approx 5600$ gives an effective statistical timing sensitivity of order $\Delta t_{\mathrm{eff}} \sim 10^{-27}$ s for 1 Hz measurement bandwidth. The precise value will depend on the actual implementation, integration time, detector stability, and residual technical noise.

The important point is that digital processing does not create information absent from the interferometer output; rather, it allows the system to approach the underlying noise floor by concentrating measurement bandwidth on the coherent gravitational signal.

### B. Sensitivity to the PPN Parameter $\gamma$

The first post-Newtonian terrestrial Shapiro delay is $\Delta t_{1PN}^{\text{Shapiro}} = 7.2 \times 10^{-20}$ s.
The corresponding statistical sensitivity to the parametrized post-Newtonian (PPN) parameter γ is approximately $\delta\gamma \approx \frac{\Delta t_{\text{eff}}}{\Delta t_{1PN}^{\text{Shapiro}}}$ where $\Delta t_{\text{eff}}$ denotes the effective post-processed timing resolution of the interferometric measurement. Using the estimated post-processed timing sensitivity, the corresponding statistical sensitivity is expected to approach $\delta\gamma \sim 10^{-7}$ .

The ultimate experimental uncertainty, however, will be determined uncertainty in Earth's gravitational parameter $GM_{earth}$, mechanical positioning accuracy; temperature gradients, polarization stability, long-term phase drift.

### C. Access to Second Post-Newtonian Effects

The predicted second post-Newtonian (2PN) contribution to the terrestrial Shapiro delay is $\Delta t_{\text{Shapiro}}^{\text{2PN}} \approx 6 \times 10^{-29}$ s, approximately nine orders of magnitude smaller than the leading first post-Newtonian (1PN) contribution.

The sensitivity estimates presented above assume a measurement bandwidth of 1 Hz and an integration time of 1 s. Because statistical uncertainty decreases as the square root of the integration time, extended data acquisition can significantly improve measurement precision. For example, continuous integration over one day (86,400 s) improves the statistical sensitivity by a factor of $\sqrt{86\,400} \approx 294$.

Additional sensitivity can be obtained by increasing the vertical separation between the fiber coils. Increasing the coil separation from 1 m to 10 m increases the gravitational propagation delay by approximately one order of magnitude, providing a corresponding improvement in sensitivity to higher-order relativistic effects.

Taken together, long integration times, improved noise suppression, and increased coil separation could extend the sensitivity of the interferometer into the regime of second post-Newtonian propagation effects. In principle, such measurements could enable experimental access to the 2PN parametrized post-Newtonian (PPN) parameters $(\beta)$ and $(\epsilon)$, complementing the primary determination of the 1PN parameter $(\gamma)$ .

### D. Device and Equipment Availability and Technology Readiness

A significant advantage of the proposed experiment is that it can be implemented entirely with commercially available photonic and electronic instrumentation, eliminating the need for custom or developmental hardware. All essential components—including long-length single-mode optical fiber, broadband light sources, semiconductor optical amplifiers, high-speed photodetectors, GHz-frequency modulators, and digital lock-in amplifiers—are mature technologies with well-established performance in precision optical measurements.

In particular, GHz-band digital lock-in amplifiers capable of coherent demodulation, high-speed digitization, and oversampling are commercially available [18,19]. For example, an 8.5 GHz digital lock-in amplifier [18] integrates a 6 GSa/s digitizer, a 14-bit analog-to-digital converter, and approximately 100 dB of dynamic reserve, providing a suitable platform for implementing the coherent signal-processing architecture described in this work.

The availability of these instruments demonstrates that the proposed experiment does not rely on speculative or emerging technologies. Rather, it leverages mature, state-of-the-art optical and electronic components whose performance has already been validated in demanding scientific and industrial applications. From an instrumentation standpoint, the proposed terrestrial measurement of the Shapiro time delay is therefore achievable with current technology.

## VIII. Tabletop Measurement of the PPN Parameter $\gamma$

The proposed experiment employs two fiber-optic spools, each containing approximately 100 km of single-mode optical fiber, arranged in the reciprocal Sagnac interferometer described previously. The measurement begins with both fiber spools positioned horizontally at the same elevation, as illustrated in Fig. 9. In this reference configuration, the two optical paths experience essentially identical gravitational potentials, establishing a zero-signal baseline. Any measured phase or propagation-time difference therefore reflects residual instrumental offsets, including thermal drifts, fiber-path asymmetries, electronic biases, polarization effects, and other common-mode disturbances.

One of the fiber spools is then translated vertically by a distance $+d$ along the local radial (nadir-pointing) direction while the second spool remains fixed, as shown in Fig. 9. This displacement places the two fiber coils at slightly different gravitational potentials, thereby producing the predicted differential Shapiro time delay while preserving the reciprocal optical geometry of the interferometer. Finally, the displaced spool is moved to the symmetric position $-d$. Because the gravitational-potential difference reverses sign, the Shapiro time delay also reverses sign, whereas the majority of instrumental offsets remain essentially unchanged. Comparing the measurements obtained at $+d$ and $-d$ therefore doubles the gravitational signal while strongly suppressing residual errors through signal reversal.

This three-position measurement protocol—consisting of the reference, positive-displacement, and negative-displacement configurations—provides a robust internal consistency check and represents a well-established differential measurement technique in precision metrology. The combination of reciprocal optical propagation, common-mode rejection, and signal reversal enables high-precision determination of the first post-Newtonian parameter (**γ**). As instrumental sensitivity and control of uncertainties continue to improve, the same experimental platform can be extended to investigate the much smaller second post-Newtonian contribution, enabling independent laboratory measurements of the higher-order PPN parameters (β) and (ε).

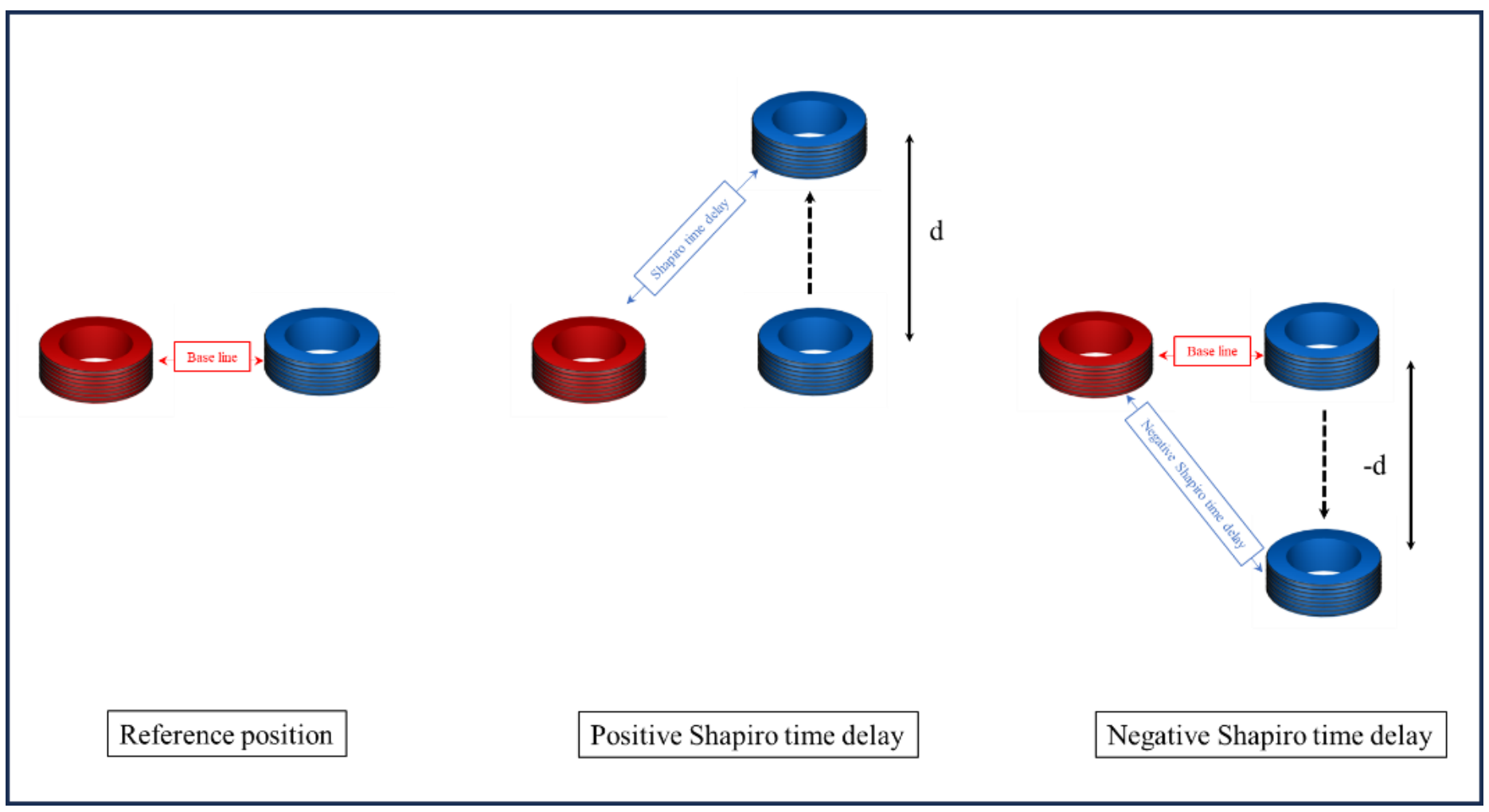

**Figure 9:** Experimental measurement sequence for determining the terrestrial Shapiro time delay. (a) Reference configuration with both fiber spools at the same elevation, establishing the instrumental zero. (b) Positive-displacement configuration (**+d**) producing a positive differential Shapiro time delay. (c) Negative-displacement configuration (**−d**) producing an equal-magnitude Shapiro time delay of opposite sign. Differencing the measurements acquired at **+d** and **−d** doubles the gravitational signal while suppressing common systematic errors.

## IX. Implications for Precision Tests of General Relativity

The proposed instrument demonstrates how the mature technology of interferometric fiber-optic gyroscopes (IFOGs) can be re-envisioned to create a new class of precision relativistic sensors. By exploiting the reciprocal Sagnac architecture to measure gravitational propagation delay rather than rotation, the same underlying technology can be adapted for laboratory determination of the parametrized post-Newtonian (PPN) parameter (**γ**) with exceptionally high intrinsic sensitivity.

A key advantage of the proposed approach is that all of the enabling technologies are already mature and commercially available. These include high-speed electro-optic modulators and photodetectors, erbium-doped fiber amplifiers (EDFAs), semiconductor optical amplifiers (SOAs), ultra-low-loss optical fibers, high-speed digital electronics, and cryogenic systems operating near 1.9 K [20,21]. Consequently, the principal challenges are no longer the availability of enabling hardware, but rather the integration of these technologies into a highly stable instrument with precise calibration, effective suppression of errors, and long-term environmental stability. From a technological standpoint, the proposed experiment therefore appears feasible using existing state-of-the-art photonic and electronic components.

A straightforward means of improving the measurement sensitivity is to increase the optical fiber length of the two Sagnac loops, thereby exploiting the approximately linear dependence of the differential Shapiro time delay on the optical propagation length. Additional sensitivity gains can be achieved by increasing the optical power incident on the photodetector, either through the use of lower-loss optical fibers, such as emerging air-core fibers, or by employing higher-power semiconductor optical amplifiers. As the received optical power increases, however, proportionally greater suppression of relative intensity noise (RIN) is required to ensure that shot noise, rather than intensity fluctuations, remains the dominant noise source. Further improvements are also expected from advances in high-speed digital signal processing. In particular, digital lock-in amplifiers with higher operating frequencies and faster analog-to-digital converters would permit operation at higher modulation frequencies, where fiber-induced noise is generally reduced, while simultaneously enabling greater oversampling and enhanced signal-to-noise ratio through coherent averaging.

Sensitivity may be enhanced even further by employing multiple Sagnac interferometers operating in parallel. This concept is analogous to the evolution of modern astronomical telescopes, where arrays of precisely phased optical elements have replaced single monolithic mirrors to achieve greater sensitivity and angular resolution. Similarly, multiple Sagnac stages, arranged either vertically or in parallel, can be configured to measure the same gravitational signal, with their outputs combined either coherently or incoherently. For statistically independent sensors, incoherent averaging improves the measurement sensitivity approximately as $\sqrt{N}$, where N is the number of interferometers. Coherent combination offers the potential for even greater improvement, provided that phase coherence is maintained across the array.

The analysis presented in this work indicates that the proposed interferometer possesses exceptionally high instrument-limited statistical sensitivity. However, as the statistical noise floor is reduced through coherent integration and digital post-processing, the ultimate experimental accuracy will increasingly be determined by systematic uncertainties rather than instrumental noise. One important limitation is the uncertainty in Earth's gravitational parameter, $GM_{earth}$, which enters directly into the theoretical prediction of the terrestrial Shapiro time delay [22]. Continued improvements in geodetic measurements of $GM_{earth}$, together with more accurate geometric calibration, environmental isolation, and long-term instrumental stability, will therefore translate directly into tighter experimental constraints on the PPN parameter $\gamma$.

For comparison, the terrestrial approach proposed by Ballmer and co-workers [5,6] is ultimately limited by the uncertainty in Newton's gravitational constant, (G), whose current relative uncertainty is approximately $10^{-5}$ [22]. In contrast, the present method is fundamentally referenced to Earth's gravitational parameter, $GM_{earth}$ whose relative uncertainty is approximately $10^{-9}$ [23]. This distinction underscores the complementary scientific value of the two approaches. Because one experiment is fundamentally referenced to G, while the other is referenced to $GM_{earth}$ , they are subject to independent uncertainties and therefore provide complementary tests of relativistic gravity. Agreement between the two would constitute a more robust validation of general relativity than either method alone.

Beyond its immediate application to measuring the first post-Newtonian parameter **γ**, the proposed interferometer establishes a versatile laboratory platform for future investigations of relativistic light propagation. As instrumental sensitivity, digital signal processing, and systematic control continue to improve, the same architecture may be extended to probe second post-Newtonian effects through measurements of the PPN parameters β and ε, thereby providing increasingly stringent laboratory tests of general relativity and alternative metric theories of gravity.

More broadly, this work illustrates how advances in fiber-optic interferometry and precision photonics can enable compact, laboratory-scale experiments that complement traditional astrophysical and space-based tests of gravitation. In this sense, the proposed instrument has the potential to transform the Shapiro time-delay experiment from a measurement traditionally requiring planetary radar observations or dedicated space missions into a repeatable laboratory experiment. Such a capability would enable measurements under controlled conditions, facilitate independent verification by multiple laboratories, and establish a scalable experimental platform for progressively more stringent tests of relativistic gravity as photonic and interferometric technologies continue to advance.

## X. Future Directions and Experimental Improvements

Continued advances in photonics and optoelectronics are expected to further enhance the capabilities of fiber-based relativistic sensors. One particularly promising direction is the development of ultra-low-loss optical fibers, including emerging air-core fiber technologies, which have the potential to reduce attenuation to approximately 0.01 dB/km. Such improvements would permit substantially longer optical interaction lengths, extending fiber-based interferometers well beyond laboratory-scale devices toward kilometer- and potentially thousand-kilometer-scale distributed sensing networks.

Another important area for future development is the suppression of relative intensity noise (RIN). Increasing the launched optical power can significantly improve interferometric sensitivity, provided that excess amplitude fluctuations are sufficiently suppressed. Cascaded semiconductor optical amplifiers (SOAs) offer one potential approach to enhanced RIN suppression, although current SOA-based techniques remain bandwidth-limited to several gigahertz. Future optical amplifier architectures and nonlinear noise-suppression methods capable of providing substantially greater RIN reduction over tens to hundreds of gigahertz would further improve the sensitivity of next-generation interferometric systems.

Continued improvements in high-speed digital instrumentation are expected to provide additional performance gains. Faster analog-to-digital converters with higher effective number of bits (ENOB), combined with improved digital signal processing, coherent demodulation, and oversampling techniques, would increase measurement precision while reducing electronic noise. Operation of these electronics at cryogenic temperatures may further enhance sensitivity by lowering thermal noise and improving overall system stability.

Taken together, these advances could ultimately enable laboratory measurements of relativistic propagation effects with sufficient precision to probe not only the first post-Newtonian (1PN) contribution but also higher-order nonlinear corrections, thereby opening a new experimental regime for terrestrial tests of general relativity.

## XI. Conclusions and Significance

The convergence of advanced fiber-optic sensing, high-speed photonics, and precision digital signal processing is creating new opportunities for laboratory tests of general relativity. We have presented a terrestrial method for measuring the gravitational Shapiro time delay using a reciprocal fiber-optic Sagnac interferometer and for determining the parametrized post-Newtonian (PPN) parameter ($\gamma$) under controlled laboratory conditions.

Unlike previous Shapiro-delay experiments based on planetary radar or spacecraft tracking, the proposed approach converts gravitational propagation delay into an optical phase measurement within a compact terrestrial interferometer. By combining long-baseline fiber optics, differential gravitational-potential sampling, signal reversal, and coherent phase detection, the architecture provides a practical framework for precision studies of relativistic light propagation.

Our analysis indicates that the instrument could achieve high intrinsic sensitivity to the first post-Newtonian Shapiro delay, with ultimate accuracy limited primarily by uncertainties such as Earth's gravitational parameter, geometric calibration, environmental stability, and long-term drift. The same platform also provides a scalable path toward measurements of second post-Newtonian effects and the associated PPN parameters ($\beta$) and ($\varepsilon$).

More broadly, this work illustrates how compact photonic interferometers can complement traditional astronomical tests of gravitation. Continued advances in low-loss fiber, noise suppression, and digital signal processing may further extend laboratory-based precision tests of spacetime physics.

If realized, this experiment would provide a new terrestrial realization of the Shapiro time-delay test, complementing the three classical solar-system tests of general relativity: the perihelion advance of Mercury, gravitational light deflection, and gravitational redshift.

The proposed experiment may also represent a new stage in the historical evolution of Shapiro time-delay measurements:

- **Planetary radar ranging (1964–1968):** Established the existence of the Shapiro delay by measuring the increased travel time of radar signals passing near the Sun.
- **Spacecraft radio science (e.g., Cassini):** Achieved substantially higher precision through deep-space radio tracking, yielding stringent constraints on the PPN parameter (\gamma).
- **VLBI and precision astrometry:** Extended relativistic tests using interferometric measurements of gravitational light deflection over astronomical baselines.
- **Coherent optical phase interferometry (this work):** Introduces a terrestrial measurement paradigm that converts gravitational propagation delay directly into an optical phase shift within a controlled laboratory interferometer.

The significance of the proposed experiment extends beyond a terrestrial demonstration of the Shapiro effect. It introduces a phase-based methodology for testing relativistic gravity, establishing coherent optical phase as a new observable for Shapiro-delay measurements alongside the traditional observables of propagation time, range, and Doppler tracking. Moreover, because the experiment is fundamentally referenced to Earth's gravitational parameter, $GM_{\text{Earth}}$, rather than Newton's gravitational constant, $G$, it possesses an independent metrological foundation and a distinct uncertainty budget. Agreement between terrestrial phase-based measurements and classical astronomical Shapiro-delay experiments would therefore provide a more stringent and robust validation of general relativity than either approach alone. While the primary objective is a laboratory determination of the first post-Newtonian PPN parameter $\gamma$, the proposed architecture is inherently scalable. With longer integration times, improved noise suppression, and increased coil separation, it could ultimately provide sensitivity to the second post-Newtonian correction to the Shapiro delay, enabling experimental constraints on the second-order coefficient $C$ and, consequently, on the combination of PPN parameters $\beta$, $\gamma$, and $\epsilon$ that governs this higher-order relativistic effect.

In this way, the proposed interferometer offers not only a new laboratory realization of the Shapiro experiment but also a scalable experimental platform for progressively more stringent tests of relativistic gravity, extending from first-order validation of the PPN parameter $\gamma$ toward future exploration of second-order post-Newtonian domain characterized by the coefficient $C$.

**Acknowledgments:** The author gratefully acknowledges Professor Irwin I. Shapiro for his insightful suggestion to extend the manuscript by including the second post-Newtonian (2PN) formulation of the Shapiro time delay. His pioneering work on the theoretical prediction and experimental confirmation of the Shapiro time delay has provided the foundation upon which the present investigation is built.

**Funding:** This research received no external funding.

**Data Availability Statement:** The original contributions presented in this study are included in the article/Supplementary Material. Further inquiries can be directed to the corresponding authors.

**Conflicts of Interest:** The authors declare no conflicts of interest.

**Authors Contributions:** Conceptualization; Methodology, Formal analysis, Writing.